# A Novel Feeder-level Microgrid Unit Commitment Algorithm Considering Cold-load Pickup, Phase Balancing, and Reconfiguration

Rongxing Hu, Ashwin Shirsat, Valliappan Muthukaruppan, Si Zhang, Yiyan Li, Lidong Song, Bei Xu, Victor Paduani, Ning Lu, *Fellow, IEEE,* Mesut Baran, *Fellow, IEEE,* Wenyuan Tang, *Member, IEEE*

*Abstract--* This paper presents a novel 2-stage microgrid unit commitment (Microgrid-UC) algorithm considering cold-load pickup (CLPU) effects, three-phase load balancing requirements, and feasible reconfiguration options. Microgrid-UC schedules the operation of switches, generators, battery energy storage systems, and demand response resources to supply 3-phase unbalanced loads in an islanded microgrid for multiple days. A performance-based CLPU model is developed to estimate additional energy needs of CLPU so that CLPU can be formulated into the traditional 2-stage UC scheduling process. A per-phase demand response budget term is added to the 1$^{st}$ stage UC objective function to meet 3-phase load unbalance limits. To reduce computational complexity in the 1$^{st}$ stage UC, we replace the spanning tree method with a feasible reconfiguration topology list method. The proposed algorithm is developed on a modified IEEE 123-bus system and tested on the real-time simulation testbed using actual load and PV data. Simulation results show that Microgrid-UC successfully accounts for CLPU, phase imbalance, and feeder reconfiguration requirements.

## I. INTRODUCTION

MICROGRIDS powered by distributed energy resources (DERs), primarily renewable generation resources and grid-forming battery energy storage systems (BESS), have attracted great interests in recent years as an effective operation mechanism to provide grid services and enhance distribution system resiliency [1].

Unit commitment (UC) is the key algorithm of the energy management system (EMSs) for scheduling generation resources in the bulk power system (BPS). However, directly applying BPS UC for microgrid EMS is oftentimes infeasible, especially for microgrids at the feeder-level. *First,* on a distribution feeder, intermittency of distributed renewables is compounded with uncertainty in loads, making combined forecasting errors much higher than those in BPSs [2]. *Second,* unlike large, synchronous generators in the BPS, DERs are constrained by both power and energy limits [3]. Particularly, the installed power and energy capacity of grid-forming BESSs or distributed generators are oftentimes insufficient to supply the microgrid load at all times in a prolonged outage that lasts for days. Therefore, demand response (DR) and feeder reconfiguration have to be frequently used to shed loads for meeting power and energy limits. *Third,* in the BPS, cold load pick-up (CLPU) [4] is seldom considered in UC. However, in an islanded microgrid, due to interruptions mainly caused by the intermittency of DERs and feeder reconfigurations, cutting off a load and resupplying it is often time required during outages, making CLPU occur more frequently. Those additional CLPU energy needs so far cannot yet be predicted by load forecasting algorithms. Note that in a distribution grid, CLPU is mainly caused by the recovery of Heating, Ventilation, and Air Conditioning (HVAC) systems, the electricity consumption of which accounts for approximately 50% of energy use in residential and office buildings [5]. *Fourth,* in BPS EMSs, loads are mostly 3-phase balanced. However, in a distribution grid, even under normally operation conditions, loads are normally unbalanced. Load imbalance can also be exacerbated by CLPU, feeder reconfiguration, or DR events. Because highly unbalanced loads can cause power quality issues, violate voltage regulation requirements, lower the sensitivity of the protection systems [6], it is critical in microgrid operation to maintain the phase imbalance within a given limit.

Thus, in this paper, we propose a novel 2-stage microgrid unit commitment (Microgrid-UC) algorithm that accounts for CLPU, using DR for three-phase load balancing, and the feasibility of reconfiguration options. Microgrid-UC manages the islanded operation of a 3-phase unbalanced distribution feeder for multiple days. The controllable resources include breakers/switches, DERs (e.g., PV farms, diesel generators, BESSs), and DR resources. The four unique considerations of Microgrid-UC are explained as follows.

**CLPU modeling:** In the literature, there are two approaches for modeling CLPU: model-based and data-driven methods. The model-based approach predicts CLPU effects by physics-based models. Using system on/off status and ambient temperatures as inputs, the electricity consumptions of HVACs are simulated to predict CLPU needs. Either exponential [7] or linearized models [8] can be used for modeling the thermodynamics process that causes CLPU. The drawback of this method is that predetermined HVAC model parameters cannot produce simulation results matching field

This research is supported by the U.S. Department of Energy's Office of Energy Efficiency and Renewable Energy (EERE) under the Solar Energy Technologies Office Award Number DE-EE0008770.

The authors are with the Department of Electrical and Computer Engineering, North Carolina State University, Raleigh, NC 27695 USA (emails: {rhu5, ashirsa, vmuthuk2, szhang56, yli257, lsong4, bxu8, vdaldeg, nlu2, baran, wtang8}@ncsu.edu).

measurements. In contrast, the data-driven approach estimates the CLPU curve parameters using historical data [9-11]. The drawback of this approach is the lack of CLPU event data. Thus, in this paper, we develop a hybrid CLPU modeling method. Instead of directly estimating the CLPU curve parameters, we use smart meter data to derive the HVAC model parameters [12]. The HVAC models can then be used to model CLPU effects for different ambient temperature and interruption durations, the results of which can be used to derive the CLPU curve parameters.

**Formulating CLPU constraints into EMS**: Conventional distribution system EMS problem formulations only account for CLPU in restoration algorithms. For example, in [13], a Mixed-Integer Linear Programming (MILP) service restoration algorithm accounts for CLPU using a linearized, delayed exponential CLPU curve. In [14], a two-block representation (one for normal loads and the other for CLPU increments) is used to eliminate the CLPU nonlinear characteristic. In [15], to capture the CLPU power after short outages, a time dependent CLPU model based on operating state evolution of thermostatically controlled loads (TCLs) is proposed. In [16], the uncertainty of CLPU, captured by the probability density functions (PDFs) of CLPU peak and duration, is included in the restoration service. However, the drawback of such formulations is the use of a predefined set of CLPU parameters, through which variations of outdoor temperature and interruption duration [17] cannot be considered. Thus, in this paper, we proposed *an adaptive CLPU estimation method* that can account for accumulated CLPU effects when picking up load groups (LGs) with different "off" durations under different ambient temperatures.

**Load unbalance:** In microgrid operation, maintaining 3-phase load balance is essential for maintaining voltage balance [18] and assuring protection relays to take correct actions [6]. In [19], the authors propose two methods for controlling distributed generations to balance 3-phase loads: adding a penalty term representing the current unbalance to the objective function and using phase power for a conservative linear approximation of the current unbalance. In [20], voltage unbalance is considered. However, using DR for mitigating unbalance in UC is an uncharted area. Thus, we *formulate a DR budget term* into the UC problem for meeting 3-phase load balancing requirements.

**Topology Scheduling:** In the literature, spanning tree (ST) [21-23] is a typical approach for distribution feeder reconfiguration. In [24], the authors analyze other radiality constraints including single-commodity flow (SCF) and the combined ST and SCF constraints. However, if we need to formulate feasibility into topology options, the complexity and runtime of the algorithm will increase drastically. In practice, because of protection settings and circuit operational limits, not all topology options are feasible under different operation conditions. Thus, we propose a *feasible topology candidate method* to ensure the feasibility of the selected topology while shortening the runtime by over 50%.

To summarize, the novelties of Microgrid-UC are three-fold. *First*, we develop an adaptive CLPU model so that CLPU can be formulated into both 24-hour ahead and intra-hour optimization problem formulations. *Second*, a DR budget term is formulated into the 24-hour ahead UC problem to balance 3-phase system loads. *Third*, we use a feasible topology candidate method to guarantee operational feasibility and reduce runtime.

The rest of this paper is organized as follows. Section II presents the proposed microgrid-UC method. Results are presented in Section III and Section IV concludes the paper.

## II. METHODOLOGY

In this section, we first introduce the layout of a typical feeder-level microgrid. Then, the assumptions, the overall framework, the problem formulation and the operational constraints of the 2-stage microgrid-UC are presented.

### A. Typical Layout of a Feeder-level Microgrid

In this paper, our focus is to develop a 2-stage microgrid-UC algorithm for managing a feeder-level microgrid by accounting for cold-load pickup, three-phase load balancing, and feeder reconfiguration. As shown in Fig. 1, a typical feeder-level microgrid is powered by a hybrid system (e.g., the MW-level PV plant collocated with a grid-forming BESS at bus 7) and supplied multiple LGs, which can be prioritized into "critical" (the red triangles) and "non-critical". The microgrid controller controls five switches (S1-S5) remotely to switch on/off LGs.

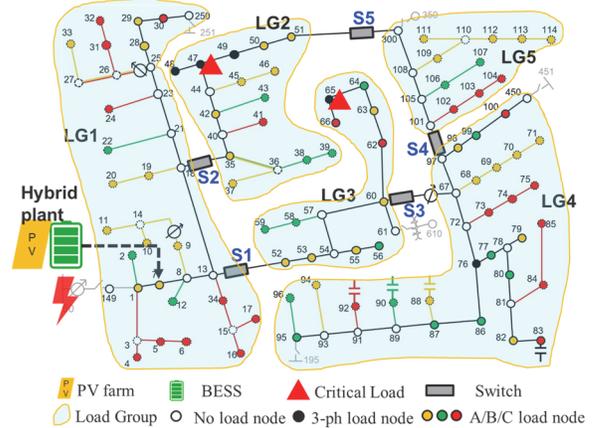

Fig. 1. An illustration of the typical layout of a feeder-level microgrid.

### B. Assumptions

We make the following assumptions regarding microgrid operation, data availability, and device controllability. *First*, the microgrid controller has access to smart meter data. The controller controls the switches and DR resources in each LG remotely via a fully functional communication network. *Second*, critical loads have their own backup generators so the goal of the microgrid controller is to reduce the use of their backup generators by weighting the critical load with higher supply priorities than the non-critical load. Only non-critical loads participate DR. *Third*, no circuit loop is allowed and there is only one grid-forming resource in the microgrid. In this paper, the BESS at bus 7 is the grid-forming resource.

### C. Scheduling Horizons and Intervals

Microgrid-UC is designed for multi-day, off-grid operations. As shown in Fig. 2, a 24-hour ahead rolling forecaster and a 30-

minute ahead forecaster are used to provide Microgrid-UC with PV, load, and weather forecasts. Figure 3 shows the scheduling horizons and intervals.

In the first stage, a rolling *24-hour ahead unit commitment* is conducted every 30-minute using 24-hour-ahead PV, load, weather forecast as inputs. The outputs are the operation schedules for the BESS, DR resources, and switches. Note that switches are switched on/off to supply/disconnect which LG for 48 30-minute scheduling intervals considering CLPU needs, phase-balancing needs, and reconfiguration.

In the second stage, a *30-minute ahead power dispatch* is conducted using 30-minute ahead PV and load forecast as inputs, while the weather input remains the same. The outputs are the operation schedules for the BESS and DR resources for six 5-minute intervals.

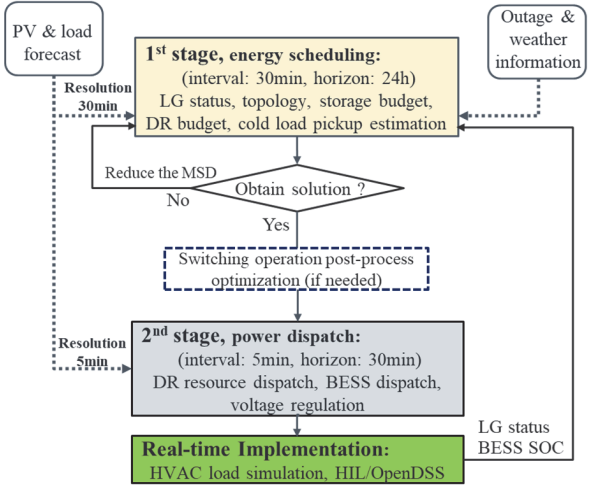

Fig. 2. The flowchart of the two-stage Microgrid-UC.

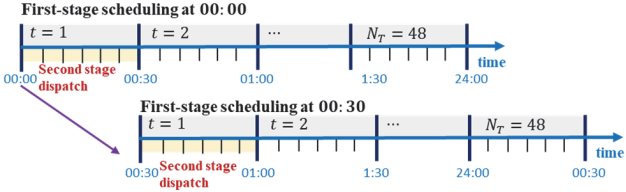

Fig. 3. Scheduling horizons and intervals of the 2-stage Microgrid-UC.

### D. Problem formulation for the 1st Stage

Let the total weighted served load be $f_1^{\text{load}}$, the total PV curtailment penalty be $f_1^{\text{PV}}$, and the total cold load pickup (CLPU) penalty be $f_1^{\text{CLPU}}$. The 1st stage objective function can be formulated as

$$\max f_1^{\text{load}} - k^{\text{PV}} f_1^{\text{PV}} - k^{\text{CLPU}} f_1^{\text{CLPU}} \quad (1)$$

$$f_1^{\text{load}} = \sum_{t=1}^{N_t} \sum_{m=1}^{N^{\text{G}}} \sum_{p \in \{a,b,c\}} U_{m,t}^{\text{G}} w_t^{\text{pref}} \left( P_{m,p,t}^{\text{Ncrit}} + w_{m,p}^{\text{crit}} P_{m,p,t}^{\text{crit}} \right) \Delta t$$
$$- k^{\text{DR}} \sum_{t=1}^{N_t} \sum_{p \in \{a,b,c\}} w_t^{\text{pref}} P_{p,t}^{\text{DR}} \Delta t \quad (2)$$

$$f_1^{\text{PV}} = 3 \sum_{t=1}^{N_t} P_{\text{PV},t}^{\text{curt}} \Delta t \quad (3)$$

$$P_{\text{PV},t}^{\text{curt}} = P_{\text{PV},t}^{\text{pred}} - P_{\text{PV},t} \quad (4)$$

where $k^{\text{PV}}$ and $k^{\text{CLPU}}$ are coefficients of the PV curtailment and the CLPU penalty; $N_t$ is the number of scheduling intervals ($N_t$ =48); $\Delta t$ is the scheduling interval ($\Delta t = 30$ minutes); $m$ is the group index, $N^{\text{G}}$ represents the total number of LGs, $p$ is the phase index; $U_{m,t}^{\text{G}}$ denotes the status of the $m^{\text{th}}$ LG (1: "on" and 0: "off"); $P_{m,p,t}^{\text{Ncrit}}$ and $P_{m,p,t}^{\text{Ncrit}}$ are the forecasted non-critical load and critical load in the $m^{\text{th}}$ LG on phase $p$ at time $t$ without considering the CLPU effect; $w_{m,p}^{\text{crit}}$ is the priority weighting of the critical load; $w_t^{\text{pref}}$ is the weighting of the customer preferred supply period[25] at time $t$; $P_{p,t}^{\text{DR}}$ is the DR budget at time $t$ on phase $p$ (note that only non-critical loads will provide DR); $P_{\text{PV},t}^{\text{curt}}$, $P_{\text{PV},t}^{\text{pred}}$ and $P_{\text{PV},t}$ are the PV curtailment, prediction and scheduled PV output on each phase.

For PV and BESS operational constraints, microgrid reserve constraints, and polygon-based linearization of active power and reactive power constraints of the inverters and the switches, please refer to [25].

### E. Minimum Service Duration Constraints

To avoid frequently switching on/off LGs, minimum service duration (MSD) constraints are introduced in [25], If an LG can be served, it is expected to be "on" for at least $D_m^{\text{MSD}}$ consecutive scheduling intervals. In this paper, we modify the formulation to allow rolling scheduling in the 1st stage considering initial service duration at the first step. As shown in Fig. 4, $\widehat{D}_{m,t}$ is the remaining service duration need, which is determined by

$$\widehat{D}_{m,t} = \max\left\{\min\left\{D_m^{\text{MSD}} - D_{m,\text{ini}}^{\text{MSD served}}, N_t - t + 1\right\}, 0\right\} \quad (5)$$

$$\sum_{z=0}^{\widehat{D}_{m,t}} U_{m,t+z}^{\text{G}} \geq \widehat{D}_{m,t}\left(U_{m,t}^{\text{G}} - U_{m,\text{ini}}^{\text{G}}\right), \ t = 1 \quad (6)$$

where $D_{m,t}^{\text{MSD served}}$ is the service duration already fulfilled in the latest $D_m^{\text{MSD}}$ steps, so $D_{m,t}^{\text{MSD served}} \leq D_m^{\text{MSD}}$. Note that MSD is treated as a soft constraint. Microgrid-UC can shut down an LG before MSD is fulfilled when there is insufficient energy supply for subsequent hours (reduce the default MSD, see Fig. 2). For example, the actual load is significantly higher or the PV is significantly lower than predicted values.

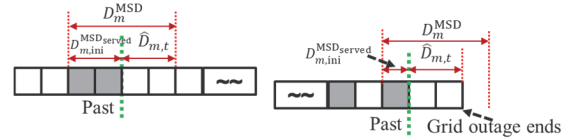

Fig. 4. The MSD requirements.

### F. CLPU Constraints

In a microgrid powered by intermittent renewable generation resources (e.g., PV and wind) and BESSs, CLPU may occur frequently. Not all LGs can be served for the entire scheduling period due to the uncertainty in generation and limitations in the BESS energy and power capacities. Thus, oftentimes, shutting down LGs is inevitable.

However, resupplying a previously "off" LG requires additional energy and power capacity to be allocated than supplying a previously "on" LG. The additional energy required in CLPU is mainly consumed by HVAC loads, which can account for approximately 50% of the total building load [5]. After an LG is turned off for a prolonged period (e.g., 60-minute), room temperature inside a building will coast out of the thermostat deadband. Thus, once the LG is switched on, all thermostatically-controlled HVAC loads will be turned on simultaneously, causing a synchronized load peak. This process can last from tens of minutes to hours depending on the "off" duration, the ambient temperature, and the thermostat setting.

To account for CLPU in Microgrid-UC, we formulate additional energy budgets required for CLPU in the 1$^{st}$ stage scheduling using a novel hybrid CLPU modeling approach, which is a major contribution to the state-of-the-art UC problem formulation.

*1) Develop the hybrid CLPU model*

To predict the CLPU effect under different ambient temperatures for different outage durations, we first need to know the HVAC model parameters. As a first step, we derive the HVAC parameters for each load profile in the Pecan Street dataset [26]. Thus, once the load profile of a load node on the feeder is selected from the Pecan Street dataset [25], the HVAC parameters for the node are known. Note that in practice, if sub-meter HVAC load profiles are not available, load disaggregation algorithms [27-28] are needed to disaggregate HVAC loads from smart meter data, after which the HVAC model parameters can be derived.

In the 123-bus test system, there are 1100 HVACs in total. Using weather forecast and LG on/off status as inputs, we can then predict the CLPU effects by modeling the HVAC consumptions for different outdoor temperatures and for different outage durations.

As shown in Figs. 5 and 6, a significant amount of additional energy beyond the "normal" consumption is needed when picking up LGs that are previously "off". After a prolonged outage, the CLPU peak is the synchronized peak of all HVAC loads, $P_m^{\text{MaxCLPU}}$. Note that if the outage occurs in a mild day, the CLPU peak may be lower than the synchronized peak (see Fig. 6, the 26 °C case). However, if a scheduling interval is 30-minute or longer, we can simplify the computation by assuming that the CLPU peak equals to $P_m^{\text{MaxCLPU}}$ regardless of how many scheduling intervals the LG has been "off".

The simulation results can be used to generate Figs. 7 and 8. As shown in Fig. 7, at a given outdoor temperature, $T_t^{\text{out}}$, the CLPU peak duration, $d_{m,t}^{\text{peak}}$, is a function of outage duration, $d_{m,t}^{\text{off}}$. The longer the outage lasts, the longer the CLPU peak duration will be. To simplify the calculation, we can linearize $d_{m,t}^{\text{peak}}$ versus $d_{m,t}^{\text{off}}$ curves so that for a given temperature, an incremental peak duration can be calculated from the slope of the curve. Note that if the maximum outage duration, $d_{m,t}^{\text{offsatu}}$, is reached, $d_{m,t}^{\text{peak}}$ is capped at $d_m^{\text{peaksatu}}$.

As shown in Fig. 8, the CLPU power decay rate, $\gamma_{m,t}^{\text{Tout}}$, is a function of outdoor temperature, $T_t^{\text{out}}$. The higher $T_t^{\text{out}}$ is, the slower the CLPU peak decays from $P_m^{\text{MaxCLPU}}$ to the steady-state HVAC consumption level, $P_{m,t}^{\text{Steady}}$. To simplify the calculation, we ignore the impact of the interruption duration on the decay rate so that an equivalent power decay rate curve can be computed with respect to $T_t^{\text{out}}$.

From those results, a performance-based CLPU model (See Fig. 9) having the following parameters can be derived: the synchronized HVAC peak load of the LG ($P_m^{\text{MaxCLPU}}$ in Fig. 6), the CLPU peak duration rate and saturation ($\tau_{m,t}^{\text{Tout}}$ and $d_m^{\text{peaksatu}}$ in Fig. 7) to get the CLPU peak duration ($d_{m,t}^{\text{peak}}$ in Fig. 9), the CLPU decay rate ($\gamma_{m,t}^{\text{Tout}}$ in Fig. 8), and the CLPU steady-state load ($P_{m,t}^{\text{Steady}}$ in Fig. 6) estimated from the outdoor temperature range in steady-state operation.

Note that we select $P_m^{\text{MaxCLPU}}$ to be the power base to make the normalized CLPU peak as 1.0 p.u. Thus, $k_{m,t}$ is the power of the HVAC load in the $m^{\text{th}}$ LG at time $t$ in per unit values with steady state value as $k_{m,t}^{\text{Steady}}$, as shown in Fig. 9.

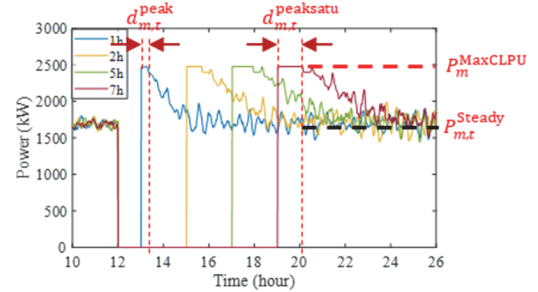

Fig. 5. CLPU effects for different interruption durations ($T^{\text{out}}$ =36 °C).

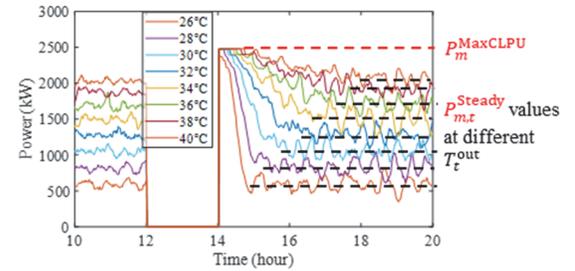

Fig. 6. CLPU effects for different outdoor temperatures (2-hour outage).

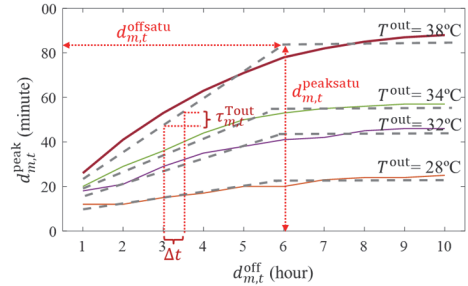

Fig. 7. CLPU effects for different outage durations and under different outdoor temperatures ($\tau_{m,t}^{\text{Tout}} = \Delta d_{m,t}^{\text{peak}}/\Delta t$, where $\Delta t = 30$ minutes).

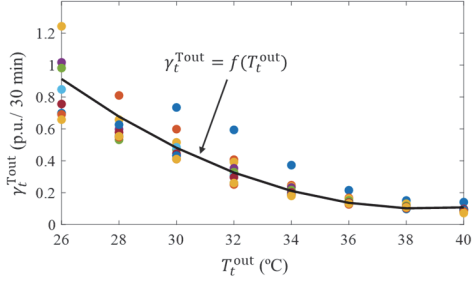

Fig. 8. CLPU decay rates with respect to outdoor temperature with dots representing simulated decay rates for different outage durations $d_{m,t}^{\text{off}}$ (Note that the curve is normalized to the synchronized CLPU peak, $P_{m,t}^{\text{MaxCLPU}}$).

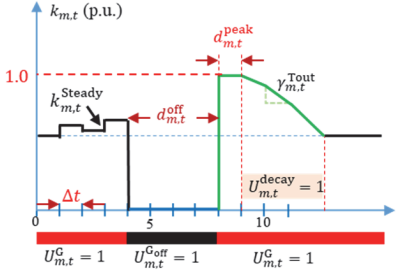

Fig. 9. Model structure of the proposed performance-based CLPU model.

*2) Estimate the CLPU peak duration*

To estimate the accumulated CLPU peak duration, we have

$$U_{m,t}^{\text{Goff}} = 1 - U_{m,t}^{\text{G}} \tag{7}$$

$$0 \leq d_{m,t} \leq MU_{m,t}^{\text{Goff}} \tag{8}$$

$$d_{m,t} + MU_{m,t}^{\text{G}} \geq d_{m,t-1} + \tau_{m,t}^{\text{Tout}}U_{m,t}^{\text{Goff}}\Delta t \tag{9}$$

$$U_m^{\text{satu}} \leq U_{m,t}^{\text{off}} \tag{10}$$

$$d_{m,t}^{\text{peak}} \geq D_{m,t}^{\text{satu}}U_m^{\text{satu}} \tag{11}$$

$$d_{m,t}^{\text{peak}} \geq d_{m,t} - MU_m^{\text{satu}} \tag{12}$$

$$d_{m,t}^{\text{re}} - d_{m,t-1}^{\text{re}} \geq d_{m,t-1}^{\text{peak}} - U_{m,t-1}^{\text{G}}\Delta t - MU_{m,t}^{\text{Goff}} \tag{13}$$

$$0 \leq d_{m,t}^{\text{re}} \leq MU_{m,t}^{\text{G}} \tag{14}$$

where $d_{m,t}$ is the estimated accumulated CLPU peak duration during interruptions without considering saturation; $U_{m,t}^{\text{Goff}}$ is the interruption status; $\tau_{m,t}^{\text{Tout}}$ is incremental CLPU duration for scheduling interval; $D_{m,t}^{\text{satu}}$ is the saturated value at time $t$ derived (Fig. 7); $M$ is a large number greater than $24 \times 60$ minutes (in our case, $M = 1500$); $U_m^{\text{satu}}$ is a binary variable indicating peak duration saturation status; $d_{m,t}^{\text{peak}}$ and $d_{m,t}^{\text{re}}$ are the estimated CLPU peak duration and the remaining peak duration, respectively.

Equation (7) determines whether the LG is "off"; (8) and (10) ensure when the LG is served, CLPU peak and CLPU peak saturation status could be 0. For each consecutive "off" interval, a resultant CLPU peak duration increment is added to the previous CLPU peak duration using (9). Note that in (9), we do not consider the saturation effect.

If the CLPU peak duration is saturated, (11) calculates the saturated CLPU peak duration and saturation status; if not, (12) ensures the accumulated CLPU peak duration by the end of the interruption duration. Thus, (12) is disabled when the CLPU peak duration is saturated. Note that the maximum CLPU peak duration is capped according to the temperature of the step (see Fig. 7).

When LGs are served intermittently, unfulfilled CLPU needs may be carried over to the next "on" cycle. To account for it, remaining CLPU peak durations can be estimated by (13) and (14). Note that minimizing the CLPU peak duration also leads to the minimization of additional energy needs for CLPU. In the results section, we will demonstrate that as a result of such considerations, Microgrid-UC tends to supply loads in consecutive intervals instead of turning them on/off frequently to minimize the total energy needed for CLPU.

*3) Set CLPU decay status*

To determine the CLPU decay status for the $m^{\text{th}}$ LG at time $t$, $U_{m,t}^{\text{decay}}$, and ensure that the CLPU decay will start only when the CLPU peak duration elapses, we add the following constraints

$$U_{m,t}^{\text{decay}} \leq U_{m,t}^{\text{G}} \tag{15}$$

$$M(1 - U_{m,t}^{\text{decay}}) \geq d_{m,t}^{\text{re}} \tag{16}$$

$$-MU_{m,t}^{\text{decay}} + M^s U_{m,t}^{\text{G}} \leq d_{m,t}^{\text{re}} \tag{17}$$

where $M^s$ is a small constant (in our case, $M^s = 0.001$).

*4) Calculate CLPU power*

After $P_{m,t}^{\text{Steady}}$ is estimated from Fig. 6 based on $T_t^{\text{out}}$, the steady state value, $k_{m,t}^{\text{Steady}}$, can be obtained by (18). The HVAC load factor $k_{m,t}$ is within peak value (1.0 p.u.) by (19).

$$k_{m,t}^{\text{steady}} = \frac{P_{m,t}^{\text{Steady}}}{P_{m,t}^{\text{MaxCLPU}}} \tag{18}$$

$$k_{m,t}^{\text{steady}} U_{m,t}^{\text{G}} \leq k_{m,t} \leq U_{m,t}^{\text{G}} \tag{19}$$

If the LG is "on" at interval $t$ and the peak decay status ($U_{m,t}^{\text{decay}}$) is still 1, the CLPU factor, $k_{m,t}^{\text{CLPU}}$, can be calculated by (20-21). We assume when a LG is turned on all HVAC loads in the LG will be turned on such that the CLPU peak is $P_{m,t}^{\text{MaxCLPU}}$, (1 p.u.), this is ensured by (20).

$$k_{m,t} - k_{m,t-1} \geq (U_{m,t}^{\text{G}} - U_{m,t-1}^{\text{G}}) - \gamma_{m,t}U_{m,t}^{\text{decay}} \tag{20}$$

$$k_{m,t}^{\text{CLPU}} = k_{m,t} - k_{m,t}^{\text{steady}}U_{m,t}^{\text{G}} \tag{21}$$

Finally, the CLPU power in kW value, $P_{m,t}^{\text{CLPU}}$, is calculated as

$$P_{m,t}^{\text{CLPU}} = k_{m,t}^{\text{CLPU}} P_{m,t}^{\text{MaxCLPU}} \tag{22}$$

*5) Formulate CLPU into the 1st stage Microgrid-UC*

To mitigate the CLPU effect, a CLPU penalty term, $f_1^{\text{CLPU}}$, consisting of three penalty factors, $k_1^{\text{CLPU}}$, $k_2^{\text{CLPU}}$, and $k_3^{\text{CLPU}}$ is added to the 1st stage Microgrid-UC problem formulation as:

$$f_1^{\text{CLPU}} = \sum_{t=1}^{N_t}\sum_{m=1}^{N^G}\left(k_1^{\text{CLPU}}P_{m,t}^{\text{CLPU}}\Delta t + k_2^{\text{CLPU}}d_{m,t}^{\text{peak}} + k_3^{\text{CLPU}}d_{m,t}^{\text{re}}\right) \tag{23}$$

$$P_{m,p,t} = P_{m,p,t}^{\text{Ncrit}} + P_{m,p,t}^{\text{crit}} \tag{24}$$

$$\sum_{l\in\Omega_m^{\text{from}}} P_{lm,p,t} = U_{m,t}^{\text{G}}P_{m,p,t} + P_{m,p,t}^{\text{CLPU}} + \sum_{n\in\Omega_m^{\text{to}}} P_{mn,p,t} \tag{25}$$

where $P_{m,p,t}$ is the baseload (i.e., the non-HVAC portion of the forecasted load) on phase $p$ in the $m^{\text{th}}$ LG; $P_{mn,p,t}$ is the active power at switch $mn$ flowing from LG $m$ to LG $n$; $\Omega_m^{\text{from}}$ and $\Omega_m^{\text{to}}$ represent the "from" LG set and "to" LG set of the $m^{\text{th}}$ LG, , respectively. Power balance within a non-source group is ensured by (25). Note that for the LG the hybrid PV plant is located at, (25) will need to consider the outputs of PV and BESS. Also, reactive power has similar.

### G. 3-Phase Load Balancing Requirements

Microgrid-UC is designed to serve 3-phase unbalanced loads. To reflect the Point of Common Coupling (PCC) power unbalance requirements, an unbalance factor $k^{\text{imb}}$ is defined as [19].

$$k^{\text{imb}} = \frac{\max(|P_{\text{pcc},p,t} - P_{\text{pcc},t}^{\text{ave}}|)}{P_{\text{pcc},t}^{\text{ave}}}, \forall p \in \{a,b,c\} \quad (26)$$

$$P_{\text{pcc},t}^{\text{ave}} = \frac{1}{3}\sum_{p\in\{a,b,c\}} P_{\text{pcc},p,t} \quad (27)$$

In a feeder-level microgrid, the load can be highly unbalanced. The current control mechanism does not allow 3-phase inverters to deliver highly unbalanced 3-phase power to the loads. Thus, it is crucial that 1-phase DR resources can be scheduled to reduce the unbalance among the 3-phase loads actively. Thus, a DR budget term is formulated in the 1st stage Microgrid-UC for balancing 3-phase loads.

$$P_{\text{pcc},p,t} = \sum_{m=1}^{N^{\text{G}}}\left(U_{m,t}^{\text{G}}P_{m,p,t} + P_{m,p,t}^{\text{CLPU}}\right) - P_{p,t}^{\text{DR}} \quad (28)$$

$$0 \le P_{p,t}^{\text{DR}} \le k^{\text{DR}} \sum_{m=1}^{N^{\text{G}}}\left(U_{m,t}^{\text{G}}P_{m,p,t} + P_{m,p,t}^{\text{CLPU}}\right) \quad (29)$$

$$-k_{\max}^{\text{imb}}P_{\text{pcc}}^{\text{ave}} \le P_{\text{pcc},p,t} - P_{\text{pcc},t}^{\text{ave}} \le k_{\max}^{\text{imb}}P_{\text{pcc}}^{\text{ave}}, \forall p \in \{a,b,c\} \quad (30)$$

where $P_{p,t}^{\text{DR}}$ denotes the DR budget of phase $p$; $P_{\text{pcc},p,t}$ and $P_{\text{pcc},t}^{\text{ave}}$ represent PCC phase power and the average power; $k^{\text{DR}}$ is the DR budget limit factor for phase balancing; $k_{\max}^{\text{imb}}$ denotes the unbalance limit.

Equation (28) is the PCC power balance equation considering the per-phase DR budget ($P_{p,t}^{\text{DR}}$) for each scheduling interval $t$; (29) limits the amount of DR resource used for phase balancing; (30) represents the 3-phase power balancing constraints at the PCC. Budgeting DR in UC for meeting 3-phase power balancing requirements is a novel consideration in UC problem formulation so we consider it as one of the main contributions of the paper.

### H. Microgrid Reconfiguration

Although Spanning tree (ST) is widely used for formulating radial topology constraints in feeder reconfiguration problems [24], there are a few drawbacks. *First*, because in the 1st stage Microgrid-UC, we won't consider voltage regulation and losses. Thus, an underlying assumption is that all topology options supplying the same LGs have the same performance. *Second*, as shown in Fig. 10, ST could assign three different topologies at time $t$, $t+1$, and $t+2$ to serve the five LGs. However, this leads to many unnecessary switching operations.

Thus, a post-process is needed to minimize the total number of switching by choosing only one topology for the three time periods (see Fig. 2). *Third*, additional constraints are needed to be added for excluding infeasible options (e.g., limited by protection settings, low voltage, or line faults). If there are many infeasible topology options, the computational complexity increases quickly. Lastly, ST does not allow multiple grid-forming resources to coordinate in a microgrid directly [25].

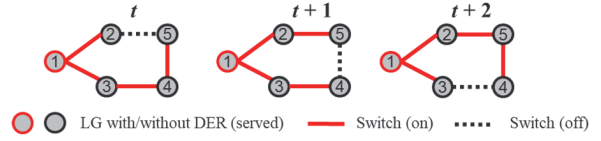

Fig. 10. Multiple feeder reconfiguration options when supplying all 5-LGs.

Thus, we propose to replace ST with searching through a list of predefined topology candidates. All candidates on the list satisfy protection settings and meet voltage regulation requirements. Thus, no post-process is needed.

The topology candidates set, $\Omega_t^{\text{topol}}$, contains $N^{\text{topol}}$ feasible topology candidates at time $t$. Let $U_{x,t}^{\text{topol}}$ be the selection status of the $x^{\text{th}}$ topology option with 1 as being selected and 0 not selected.

Because only one topology candidate can be selected at each interval, we have

$$\sum_{x=1}^{N^{\text{topol}}} U_{x,t}^{\text{topol}} = 1 \quad (31)$$

Then, the LG on/off status is determined by

$$\left[U_{1,t}^{\text{G}}, U_{2,t}^{\text{G}}, \ldots U_{N^{\text{G}},t}^{\text{G}}\right]^{\text{T}} = \mathcal{M}_t^{\text{G}}\left[U_{1,t}^{\text{topol}} U_{2,t}^{\text{topol}} \ldots U_{N^{\text{topol}},t}^{\text{topol}}\right]^{\text{T}} \quad (32)$$

where $\mathcal{M}_t^{\text{G}}$ is the $N^{\text{G}} \times N^{\text{topol}}$ LG mapping matrix.

Finally, the switch status is determined by

$$\left[U_{1,t}^{\text{SW}} U_{2,t}^{\text{SW}} \ldots U_{N^{\text{SW}},t}^{\text{SW}}\right]^{\text{T}} = \mathcal{M}_t^{\text{SW}}\left[U_{1,t}^{\text{topol}} U_{2,t}^{\text{topol}} \ldots U_{N^{\text{topol}},t}^{\text{topol}}\right]^{\text{T}} \quad (33)$$

where $\mathcal{M}_t^{\text{SW}}$ is the $N^{\text{SW}} \times N^{\text{topol}}$ switch status mapping matrix.

### I. The 2nd Stage Microgrid-UC Problem Formulation

The objective function of the 2nd stage Microgrid-UC minimizes the amount of DR usage, $f_2^{\text{DR}}$, the PV curtailment, $f_2^{\text{PV}}$, and the BESS energy deviation from its budget, $f_2^{\text{BESS}}$.

$$\min f_2^{\text{DR}} + k_2^{\text{BESS}}f_2^{\text{BESS}} + k_2^{\text{PV}}f_2^{\text{PV}} \quad (34)$$

$$f_2^{\text{DR}} = \sum_{t'=1}^{T_2}\sum_{m=1}^{N^{\text{G}}}\sum_{i=1}^{N_m^{\text{node}}}\sum_{p\in\{a,b,c\}} U_{m,i,p,t'}^{\text{DR}} P_{m,i,p,t'} \Delta t' \quad (35)$$

$$P_{m,i,p,t'} = U_{m,i,p,t'}^{\text{DR}}\left(P_{m,i,p,t'} + k_{m,t'}^{\text{CLPU}}P_{m,i,p,t'}^{\text{MaxCLPU}}\right) \quad (36)$$

$$f_2^{\text{BESS}} = \Delta E_2^- \quad (37)$$

$$\Delta E_2^+ - \Delta E_2^- = E_{2,T_2} - E_{1,t} \quad (38)$$

$$f_2^{\text{PV}} = 3\sum_{t'=1}^{T_2} P_{\text{pv},t'}\Delta t' \quad (39)$$

where $k_2^{\text{BESS}}$ and $k_2^{\text{PV}}$ are the weight coefficients of the BESS energy storage deviation and the PV curtailment; $\Delta t'$ is the scheduling interval (5 minutes); $T_2$ is the scheduling horizon

(30 minutes) of the 2$^{nd}$ stage; $N_m^{node}$ is the number of nodes in LG $m$, $i$ is the node index; $U_{m,i,p,t'}^{DR}$ is the demand response actions (1: turn off the load); $k_{m,t'}^{CLPU}$ is the CLPU factor calculated by interpolating the first stage's estimated CLPU factor $k_{m,t}^{CLPU}$ linearly; $E_{1,t}$ is the BESS energy setpoint for the 2$^{nd}$ stage (it is the BESS energy by the end of step $t$ from the first stage optimization); $E_{2,T_2}$ is the BESS energy by the end of the last step $T_2$ at the 2$^{nd}$ stage; $\Delta E_2^+$ and $\Delta E_2^-$ are the positive and negative energy storage deviations (both are non-negative).

The CLPU effect is included by (36) in the 2$^{nd}$ stage, (37) only penalizes the negative BESS deviation and (38) calculates the BESS energy deviations.

In the 2$^{nd}$ stage, we perform linear power flow [29] with voltage regulation constraints, the hybrid PV plant voltage is set at 1.03 p.u., for other nodes, the voltage should be regulated within [0.95,1.05] p.u.[25]. The DR is dispatched to meet power balance, and 3-phase power balancing requirements and system reserve limits. The PV and BESS operational constraints, microgrid reserve constraints, and polygon-based linearization of active power and reactive power constraints of the inverters and the switches are similar to those in the 1$^{st}$ stage, the imbalance limit (30) is also included.

## III. SIMULATION RESULTS

In this paper, the performance of the proposed algorithm is demonstrated on a modified IEEE-123 bus system, as shown in Fig. 1. If the main grid power is lost, the feeder will be supplied by a hybrid PV plant (on node 7) consisting of a PV farm (rated at 4.2 MW) and a BESS (rated at 3 MW/6 MWh).

The PV forecast data are obtained based on actual measurements of a 5MW PV farm. As shown in Fig. 11, there are significant PV forecast errors on the second day. The charging/discharging efficiency of the BESS is 95%. The minimum and maximum SOC of the BESS are set at 20% and 90%, respectively, and the initial SOC is 90%. As the only grid-forming source, the BESS regulates the system voltage at 1.03 p.u..

Nodes 47 and 65 have critical loads with the load priority weighting ($w_{m,p}^{crit}$) set as 2. The outage length is 48 hours. There are two customer preferred service periods: 7:00-9:00 a.m. and 18:00-20:00 p.m.. The preferred time weighting ($w_t^{pref}$) is 1.5.

The load profiles at each load node in the 123-bus system are populated using the Pecan Street dataset [26] using methods introduced in [25]. As shown in Figs. 12 and 13, the 3-phase loads in an LG can be very unbalanced (e.g. LG1). The non-HVAC load (i.e., the base load) is forecasted using the method introduced in [30]. By combining the baseload and the modeled HVAC loads, we can calculate the total house level loads. CLPU parameters are obtained by the method presented in Section II.F, we assume each LG has the same CLPU parameters except $P_{m,t}^{Steady}$. The corresponding weather data are downloaded from NOAA.

The proposed Microgrid-UC algorithm has been formulated as an MILP problem, it is solved by the CPLEX 12.10 integrated with MATLAB 2019b on a desktop with I9-9900 CPU and 64G RAM. The EMS-OpenDSS simulation is conducted using the Matlab COM interface.

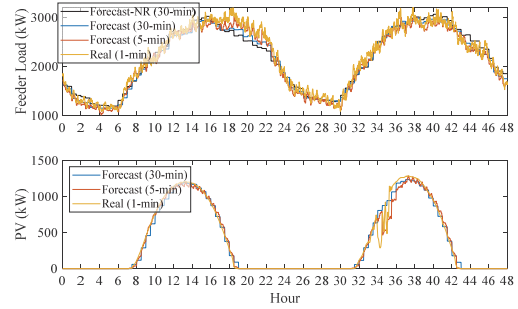

Fig. 11. Total load and per-phase PV profiles (forecasted and real) on the 123-bus feeder. Forecast-NR denotes the non-rolling day-ahead forecast.

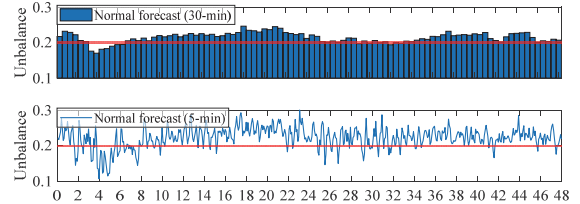

Fig. 12. Unbalance factor of the total feeder load.

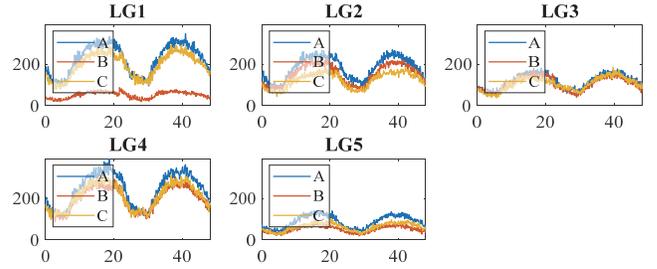

Fig. 13. Per-phase load forecast (5-min) for the five LGs.

### A. Cold Load Pickup

Two cases (with and without CLPU estimation) are set up to quantify the impact of CLPU on microgrid EMS. To show the CLPU effect more clearly, the phase imbalance limits are ignored in the 2 stages and the DR budget is not considered in the 1$^{st}$ stage UC. The reconfiguration is constrained by ST. We run the EMS-OpenDSS co-simulation using baseload forecast and the simulated HVAC load, both of which are 1-minute data.

The results are presented in Figs. 14-19. In the figures, "RT" denotes real-time simulation results and "EMS" denotes the 2$^{nd}$ stage Microgrid-UC dispatching results. From the results, we have made the following observations:

- As shown in Fig. 14 (a), if we don't consider CLPU, MSD will be frequently violated and there will be many short interruptions (e.g., nine 30-minute interruptions). This is caused by frequent overspending of the given energy budget when picking up previously "off" LGs. In contrast, when CLPU is considered (see Fig. 15 (a)), MSD is satisfied most of the time. The minimum service duration is 1 hour, but it occurred only once.
- As shown in Fig. 14 (b), without the CLPU estimation, the actually served load can be over two times of the Microgrid-UC dispatch values, showing that significant amount of additional energy is needed to meet CLPU when picking up those "off" groups. This leads to a deficiency in energy budgeted for subsequent hours.

Thus, LGs are frequent turned off for compensating the deficiency. As shown in Fig. 15 (b), by taking CLPU into consideration, the served load matches the actual load closely. This shows that Microgrid-UC optimizes the supply sequence of LGs to minimize CLPU effects.

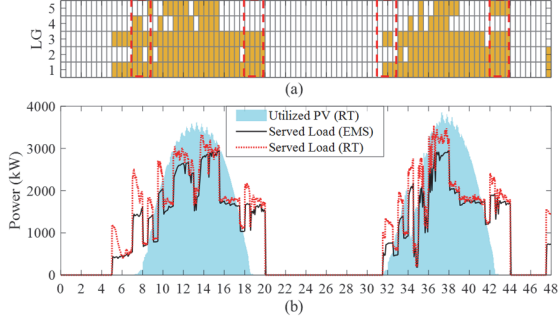

Fig. 14. Without CLPU estimation: (a) LG status with "yellow" as served, (b) served load profiles.

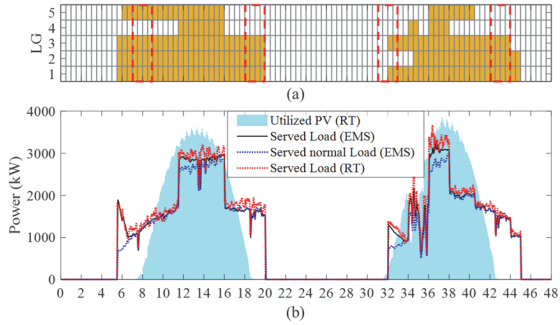

Fig. 15. With CLPU estimation: (a) LG status with "yellow" as served, (b) served load profiles.

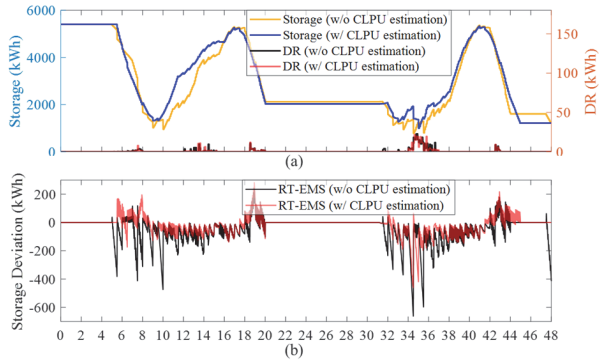

Fig. 16. (a) BESS and DR power dispatched by Microgrid-UC, (b) BESS deviation between the real-time simulation results and the dispatch results.

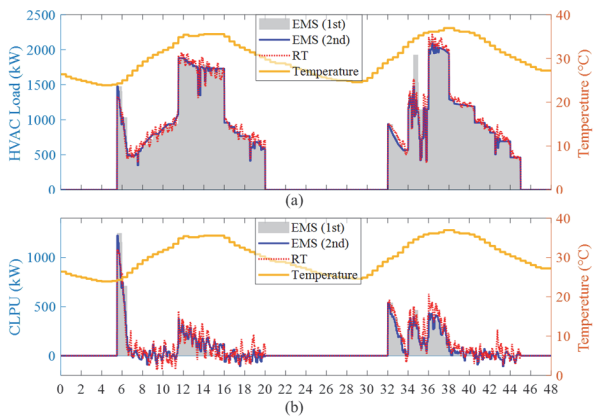

Fig. 17. With CLPU estimation: estimated and simulated (a) HVAC load, (b) CLPU effect.

- As shown in Fig. 16 (b), without considering CLPU, the actual BESS storage frequently drops below the dispatched value, causing suboptimal solutions, violation of MSD constraints, and more interruptions. When considering CLPU (see Fig. 17), even if CLPU lasts for hours, Microgrid-UC can still capture the CLPU magnitude and duration very well.
- We do observe that the DR dispatching (see Fig. 16 (a)), mainly caused by the forecast error, may cause more CLPU estimation errors because the load shedding action can also lead to CLPU (see Fig. 17). The DR caused CLPU has not yet been account for in the proposed algorithm.
- Figures 18 and 19 show that phase unbalance and voltage fluctuations are more severe when CLPU is not considered. This is because the unbalance can be exacerbated by the additional CLPU energy needs.
- As shown in Table I, considering CLPU in the Microgrid-UC can serve more loads and provide better service to critical loads while meeting MSD requirements and mitigating CLPU effects.

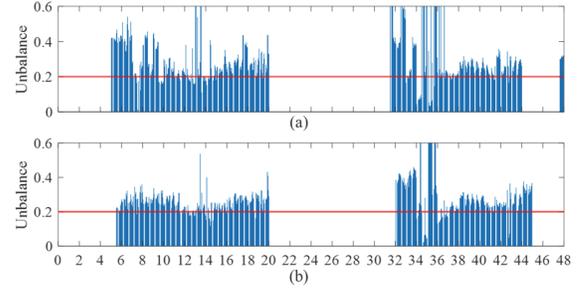

Fig. 18. Phase unbalance in real-time implementation: (a) without CLPU estimation, (b) with CLPU estimation.

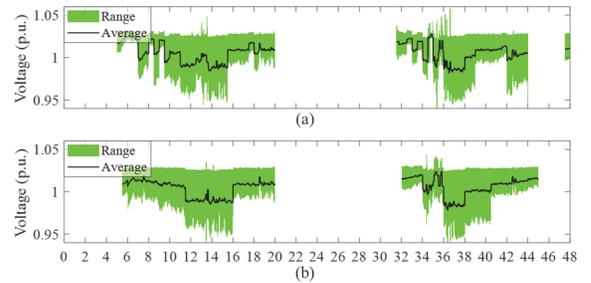

Fig. 19. Load node voltages in real-time implementation: (a) without CLPU estimation, (b) with CLPU estimation.

TABLE I
MICROGRID PERFORMANCE COMPARISON WITH AND WITHOUT CLPU

| CLPU estimation | Served energy (kWh) | Served energy during preferred periods (kWh) | Critical load served time (h) [Node 47 + Node 65] | Served HVAC energy (kWh) | CLPU (kWh) |
|---|---|---|---|---|---|
| without | 52913 | **12284** | 22.5+26.5 | **30682** | 5277 |
| with | **53271** | 10420 | **25+26.5** | 30104 | **3711** |

### B. DR Budget

In this section, the benefit of budgeting DR for balancing the 3-phase load in the first stage UC is quantified. As shown in Table II, we set up six cases with different DR budget limits and 3-phase load imbalance requirements. Note that we only run the Microgrid-UC in standalone and CLPU is considered for all 6 cases.

TABLE II
MICROGRID PERFORMANCE COMPARISON CONSIDERING DR BUDGETS

| Case | Unbalance limits | 1st stage | | | | 2nd stage | | |
|---|---|---|---|---|---|---|---|---|
| | | DR budget (% served load) | Served energy (kWh) | DR budget (kWh) | Served energy (preferred period) (kWh) | Critical load served time (h) | Served energy (kWh) | Demand Response (kWh) | PV Curtailment (kWh) |
| 1 | - | 0 | 56264 | 0 | 9966 | 26.5+26.5 | 54020 | 1963 | 127 |
| 2 | - | 10% | 55752 | 1503 | 10045 | 26+29 | 54191 | 2767 | 34 |
| 3 | 0.2 | 0 | 21677 | 0 | 0 | 8.5+8.5 | 21505 | 80 | 29234 |
| 4 | 0.2 | 10% | **55811** | 2101 | 9237 | 25.5+26 | **54164** | 3433 | **19** |
| 5 | 0.2 | 20% | 54999 | 3923 | 10226 | **28+28** | 53789 | 4806 | 420 |
| 6 | 0.2 | 30% | 55177 | 6754 | **11013** | **28+28** | 54119 | 7509 | 100 |

The following observations are made:
- If phase imbalance is not considered (Cases 1 and 2), the DR budget mainly prolongs the service time of the critical load, increases the served load during preferred periods, and reduces PV curtailments.
- If phase imbalance is considered in both stages (Case 3) without DR budget considerations in the 1st stage, the microgrid serves the least load, curtails the most PV, results in the least service time for critical loads, and serves the smallest amount load in preferred periods.
- If phase imbalance is considered in both stages, with the DR budget in 1st stage at 10% (case 4), there are more load severed with the least PV curtailment. However, further increasing the DR budget limit from 10% to 30% (cases 5 and 6) will not lead to system performance improvement. This indicates that recruiting 10% load for DR for each scheduling interval is the optimal operation setting for maintaining phase balance, serving critical loads, and reducing PV curtailments.
- Note that the flexibility of microgrid service is topology specific. In our case, the limiting factor is the imbalance load in LG1 because the hybrid PV plant is located in LG1. To compensate for the imbalance in LG1, Microgrid-UC may give priority to serve LGs that can balance the LG1 loads. By scheduling DR to remove the unbalance in each LG, we not only enable the microgrid to provide service in more hours (see Fig. 20), but also improves the fairness and flexibility when serving LGs.
- Figure 21 shows how budgeting DR facilitates the unbalance control in case 4. Note that there are a few intervals when all three phase DR resources are used for achieving other benefits (like the improvements of case 2 compared to case 1).

We conduct EMS-OpenDSS co-simulation for case 4 to verify the performance of the unbalance control. As shown in Fig. 22, the unbalance level is significantly lower than that of the case without unbalance limit (see Fig. 18). Note that the unbalance factor can be slightly over 0.2 due to forecast errors.

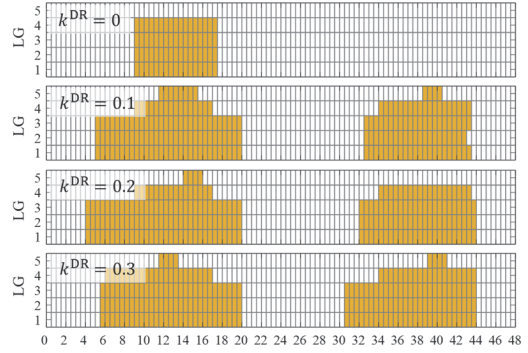

Fig. 20. Service status of each LG with different DR budget settings ($k^{DR}$).

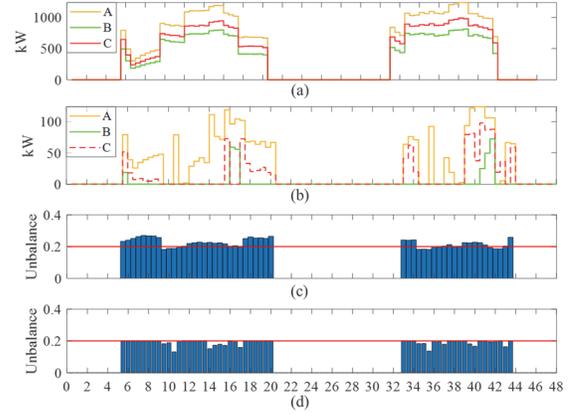

Fig. 21. PCC power scheduling (1st stage) of case 4. (a) PCC load forecast (no DR), (b) Scheduled DR actions, (c) unbalance factors before DR, and (d) unbalance factors after DR.

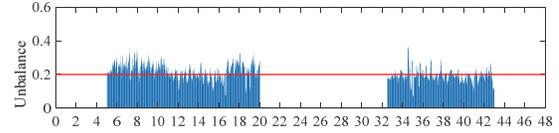

Fig. 22. Phase unbalance in real-time implementation (case 4).

*C. Topology Scheduling*

The 5-LG system has 16 topology options (see Fig. 23). To compare the topology candidate method with the ST method, we set up four cases: **ST** (spanning tree), **Candi-all** (all 16 options), **Candi-R4** (omit options 13, 14, 16), and **Candi-R5** (omit options 8, 13, 14, 16). Note that options 13, 14, and 16 are omitted because, when all 5 LGs are served, option 15 is the feeder's default topology (better protection settings). Option 8 is not selected because it leads to voltage regulation issues due to heavy loading. The DR budget limit is 10% in stage 1 and the unbalance limit is 0.2. CLPU is considered in all cases.

The service performances of the four cases are the same. The topology scheduling mainly impacts the runtime of the 1st stage UC (the 2nd stage only needs less than 5 seconds). Removing infeasible/problematic topology candidates from the searching list has two benefits: simplifying the protection settings and shortening the Microgrid-UC runtime.

As shown in Fig. 24, the maximum runtime of *Candi-all* is the highest, followed by *ST*. *Candi-R4* and *Candi-R5* reduce calculation time significantly, making the runtime for each scheduling within 80s. In practice, to serve the same LGs, there may be only 1 path that is feasible due to protection settings or for meeting service requirements. Therefore, using a list of candidates will greatly simplify the UC computing time while avoiding infeasible solutions.

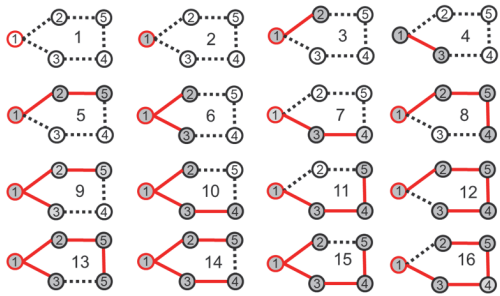

Fig. 23. Sixteen microgrid topologies. Each circle represents an LG, a grey filled circle denotes a served LG. Each line represents a switchable branch. Red solid lines as "on" branches and black dotted lines are "off".

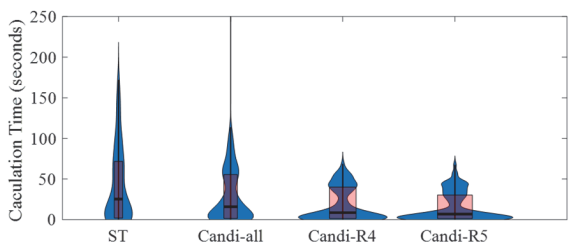

Fig. 24. 1st stage UC runtime comparison.

## IV. Conclusions

This paper proposed a 2-stage Microgrid-UC method for supplying 3-phase unbalanced loads in an islanded microgrid for multiple days. Microgrid-UC uses an adaptive CLPU estimation method to schedule for CLPU, which can account for temperature variations when interruption durations vary. Microgrid-UC balances the 3-phase loads to improve microgrid service flexibility by budgeting DR for load balancing services in the first stage. Lastly, Microgrid-UC used topology candidate based reconfiguration method to improve the computational efficiency. The superior performance of Microgrid-UC compared with the conventional UC algorithms are verified by the test cases. In our follow-up paper, we plan to extend the Microgrid-UC algorithm for managing microgrids with multiple grid-forming resources and across multiple feeders.

## V. References

[1] A. Hussain, V.-H. Bui, and H.-M. Kim, "Microgrids as a resilience resource and strategies used by microgrids for enhancing resilience," *Appl. Energy*, vol. 240, pp. 56–72, Apr. 2019.
[2] H. Jiang, Y. Zhang, E. Muljadi, J. J. Zhang, and D. W. Gao, "A short-term and high-resolution distribution system load forecasting approach using support vector regression with hybrid parameters optimization," *IEEE Trans. Smart Grid*, vol. 9, no. 4, pp. 3341–3350, Jul. 2018.
[3] H. A. Rahman, *et al.*, "Operation and control strategies of Integrated Distributed Energy Resources: A Review," *Renewable and Sustain. Energy Rev.*, vol. 51, pp. 1412–1420, 2015.
[4] E. Agneholm and J. Daalder, "Cold load pick-up of residential load," Proc. *Inst. Elect. Eng., Gen. Transm. Distrib*., vol. 147, no. 1, pp. 44-50, Jan. 2000.
[5] U.S. Energy Information Administration (2013, Mar. 11). Heating and cooling no longer majority of U.S. home energy use. [Online]. Available: https://www.eia.gov/todayinenergy/detail.php?id=10271.
[6] M. W. Davis, R. Broadwater, and J. Hambrick, "Modeling and testing of unbalanced loading and voltage regulation: Final report," Nat. Renew. Energy Lab., U.S. Dept. Energy, Golden, CO, USA, Tech. Rep. NREL/SR-581-41805, Jul. 2007.
[7] W. Lang, M. Anderson, and D. Fannin, "An analytical method for quantifying the electrical space heating component of a cold load pick up," *IEEE Trans. Power App. Syst.*, vol. PAS-101, no. 4, pp. 924–932, Apr. 1982.
[8] J. Aubin, R. Bergeron, and R. Morin, "Distribution transformer overloading capability under cold-load pickup conditions," *IEEE Trans. Power Del.*, vol. 5, no. 4, pp. 1883–1891, Oct. 1990.
[9] M. Gilvanejad, H. Askarian Abyaneh, and K. Mazlumi, "Estimation of cold-load pickup occurrence rate in distribution systems," *IEEE Trans. Power Del.*, vol. 28, no. 2, pp. 1138–1147, Apr. 2013.
[10] F. Bu, K. Dehghanpour, Z. Wang, and Y. Yuan, "A data-driven framework for assessing cold load pick-up demand in service restoration," *IEEE Trans. Power Syst.*, vol. 34, no. 6, pp. 4739–4750, Nov. 2019.
[11] A. Khurram, R. Malhamé, L. Duffaut Espinosa, and M. Almassalkhi, "Identification of hot water end-use process of electric water heaters from energy measurements," *Elec. Power Syst. Res.,* vol. 189, p. 106625, 2020.
[12] J. Lu, "Modeling and controller design of a community microgrid," Ph.D. dissertation, Dept. Elect. Eng., North Carolina State Univ., Raleigh, NC, USA, 2018
[13] B. Chen, C. Chen, J. Wang, and K. L. Butler-Purry, "Multi-time step service restoration for advanced distribution systems and Microgrids," *IEEE Trans. Smart Grid*, vol. 9, no. 6, pp. 6793–6805, Nov. 2018.
[14] A. Arif, *et al*., "Optimizing service restoration in distribution systems with uncertain repair time and demand," *IEEE Trans. Power Syst*., vol. 33, no. 6, pp. 6828–6838, Nov. 2018.
[15] M. Song, R. R. nejad, and W. Sun, "Robust distribution system load restoration with time-dependent cold load pickup," *IEEE Trans. Power Syst.*, vol. 36, no. 4, pp. 3204-3215, Jul. 2021.
[16] Y. L. Li, *et al*., "Restoration strategy for active distribution systems considering endogenous uncertainty in cold load pickup," *IEEE Trans. Smart Grid*, vol. 13, no. 4, pp. 2690–2702, Jul. 2022.
[17] J. McDonald and A. Bruning, "Cold load pickup," *IEEE Trans. Power App. Syst.*, vol. PAS-98, no. 4, pp. 1384-1386, 1979.
[18] Y. Du, X. Lu, H. Tu, J. Wang, and S. Lukic, "Dynamic microgrids with self-organized grid-forming inverters in unbalanced distribution feeders," *IEEE J. Emerg. Sel. Topics Power Electron*., vol. 8, no. 2, pp. 1097-1107, Jun. 2020.
[19] Z. Wang, J. Wang, and C. Chen, "A three-phase microgrid restoration model considering unbalanced operation of distributed generation," *IEEE Trans. Smart Grid*, vol. 9, no. 4, pp. 3594-3604, Jul. 2018.
[20] M. Roustaee and A. Kazemi, "Multi-objective Energy Management Strategy of unbalanced multi-microgrids considering technical and economic situations," *Sustain. Energy Technol. Assessments*, vol. 47, Oct. 2021.
[21] J. Li, X.-Y. Ma, C.-C. Liu, and K. P. Schneider, "Distribution system restoration with microgrids using spanning tree search," *IEEE Trans. Power Syst.*, vol. 29, no. 6, pp. 3021-3029, Nov. 2014.
[22] R. A. Jabr, R. Singh, and B. C. Pal, "Minimum loss network reconfiguration using mixed-integer convex programming," *IEEE Trans. Power Syst.*, vol. 27, no. 2, pp. 1106-1115, May 2012.
[23] X. Huang, Y. Yang, and G. A. Taylor, "Service restoration of distribution systems under distributed generation scenarios," *CSEE J. Power Energy Syst.*, vol. 2, no. 3, pp. 43-50, Sep. 2016.
[24] Y. Wang, Y. Xu, J. Li, J. He, and X. Wang, "On the radiality constraints for distribution system restoration and Reconfiguration Problems," *IEEE Trans. Power Syst*., vol. 35, no. 4, pp. 3294-3296, Jul. 2020.
[25] R. Hu *et al*., "A Load Switching Group based Feeder-level Microgrid Energy Management Algorithm for Service Restoration in Power Distribution System," 2021 IEEE Power & Energy Society General Meeting (PESGM), Washington, DC, USA, 2021, pp. 1-5.
[26] Pecan Street dataset. (Oct. 2020) [online]: http://www.pecanstreet.org/
[27] M. Liang, *et al*., "HVAC load Disaggregation using Low-resolution Smart Meter Data," 2019 IEEE Power & Energy Society Innovative Smart Grid Technologies Conference (ISGT), Washington, DC, USA, 2019, pp. 1-5.
[28] H. Kim, *et al*., "An ICA-based HVAC load disaggregation method using Smart Meter Data," 2022, *arXiv: 2209.09165*.
[29] O. Bassey, C. Chen, and K. L. Butler-Purry, "Linear power flow formulations and optimal operation of three-phase autonomous droop-controlled microgrids," *Elec. Power Syst. Res.,* vol. 196, p. 107231, Apr. 2021.
[30] Y. Li, S. Zhang, R. Hu, and N. Lu, "A meta-learning based distribution system load forecasting model selection framework," *Appl. Energy*, vol. 294, Jul. 2021.